\documentclass[aip,amsmath,amssymb,reprint]{revtex4-1}

\usepackage{graphicx}
\usepackage{dcolumn}
\usepackage{bm}
\usepackage[utf8]{inputenc}
\usepackage[T1]{fontenc}
\usepackage{mathptmx}
\usepackage{url}
\usepackage{color}
\usepackage{upgreek}
\usepackage{verbatim}

\definecolor{darkred}{rgb}{0.5, 0, 0}
\definecolor{darkgreen}{rgb}{0, 0.5, 0}
\definecolor{darkblue}{rgb}{0.1, 0.1, 0.7}

\usepackage[bookmarks=true,
            bookmarksnumbered=false,
            bookmarksopen=true,
            breaklinks=true,
            pdfborder={0 0 0},
            final=true,
            backref=none,
            colorlinks=true,
            linkcolor=darkblue,
            menucolor=darkblue,
            citecolor=darkblue,
            urlcolor=darkblue]{hyperref}

\newcommand{\micro}{${\upmu}$}

\newcommand{\um}{$\,$\micro m}

\newcommand{\revision}[1]{#1}

\begin{document}

\preprint{AIP/123-QED}

\title{Exciton-polaritons in GaAs-based slab waveguide photonic crystals}

\author{C. E. Whittaker}%
\affiliation{Department of Physics and Astronomy, University of Sheffield, Sheffield S3 7RH, United Kingdom}%
\author{T. Isoniemi}
\affiliation{Department of Physics and Astronomy, University of Sheffield, Sheffield S3 7RH, United Kingdom}%
\author{S. Lovett}%
\affiliation{Department of Physics and Astronomy, University of Sheffield, Sheffield S3 7RH, United Kingdom}%
\author{P. M. Walker}%
\affiliation{Department of Physics and Astronomy, University of Sheffield, Sheffield S3 7RH, United Kingdom}%
\author{S. Kolodny}
\affiliation{ITMO University, St. Petersburg 197101, Russia}%
\author{V. Kozin}
\affiliation{ITMO University, St. Petersburg 197101, Russia}%
\author{I. V. Iorsh}
\affiliation{ITMO University, St. Petersburg 197101, Russia}%
\author{I. Farrer}
\affiliation{Department of Electronic and Electrical Engineering, University of Sheffield, S3 7HQ United Kingdom}%
\author{D. A. Ritchie}
\affiliation{Cavendish Laboratory, University of Cambridge, CB3 0HE Cambridge, UK}%
\author{M. S. Skolnick}%
\affiliation{Department of Physics and Astronomy, University of Sheffield, Sheffield S3 7RH, United Kingdom}%
\author{D. N. Krizhanovskii}
\email{d.krizhanovskii@sheffield.ac.uk}
\affiliation{Department of Physics and Astronomy, University of Sheffield, Sheffield S3 7RH, United Kingdom}%
\affiliation{ITMO University, St. Petersburg 197101, Russia}%


\begin{abstract}
We report the observation of band gaps for \revision{low loss} exciton-polaritons propagating \revision{outside the light cone} in GaAs-based planar waveguides patterned into two-dimensional photonic crystals. By etching square lattice arrays of shallow holes into the uppermost layer of our structure, we open gaps on the order of 10 meV in the photonic mode dispersion, whose size and light-matter composition can be tuned by proximity to the strongly coupled exciton resonance. We demonstrate gaps ranging from almost fully photonic to highly excitonic. Opening a gap in the exciton-dominated part of the polariton spectrum is a promising first step towards the realization of quantum-Hall-like states arising from topologically nontrivial hybridization of excitons and photons. 
\end{abstract}

\maketitle


The hybridization of photons and quantum well excitons leads to the formation of exciton-polaritons, quasiparticles combining high-speed propagation with large nonlinearity and susceptibility to magnetic fields. These favourable properties arising from their mixed light-matter nature have made polaritons highly attractive candidates for novel semiconductor-based optical devices incorporating nonlinearity \cite{Sanvitto2016} and robustness against disorder induced by topology \cite{AMO2016934}. In order to direct and manipulate the flow of polaritons accordingly, appropriate tailoring of the potential landscape is required \cite{Schneider_2016}. In Bragg microcavities, the most mature platform for polariton research to date, lithographic patterning has been used to strongly modify the photonic mode dispersion resulting in gapped polariton band structures enabling coherent devices \cite{PhysRevLett.109.216404,PhysRevLett.110.236601,Sturm2014,doi:10.1063/1.4936158} and topological lasers \cite{St-Jean2017,Klembt2018}.

An alternative geometry to conventional microcavities for the study of polaritons is slab waveguides (WGs), in which a guided electromagnetic mode confined by total internal reflection strongly couples to quantum well excitons \cite{doi:10.1063/1.4773590}. This configuration not only has relative ease of fabrication, but also far greater suitability for integration into on-chip circuits owing to the large in-plane propagation velocities. Coherent \cite{Jamadi2018,Suarez-Forero:20} and continuum \cite{Walker2019,2009.02059} light sources have already been demonstrated using this ``horizontal'' geometry, \revision{and the thin layer structure facilitates enhancement of nonlinearities using dipolar polaritons~\cite{Rosenberg2018,SuarezForero2021}.} 

\revision{Similar to microcavities the potential landscape can be engineered to emulate novel physical systems, but WGs can achieve this without the need to etch several micrometers of material~\cite{PhysRevB.80.201308}. A full quantum theory of strong light-matter coupling in photonic crystal (PhC) slabs with embedded quantum wells was given by Gerace and Andreani over a decade ago~\cite{Gerace2007}. Experimentally, PhCs and bound states in the continuum have been demonstrated using monolayer semiconductors (MS) placed on top of periodically patterned dielectric WGs~\cite{Zhang2018,Chen2020,Kravtsov2020}. PhC were also made using pillars of hybrid organic-inorganic perovskites embedded in a homogeneous dielectric~\cite{Dang2020}. In the area of topologically non-trivial polaritons, propagating edge states protected by breaking of a pseudo-time-reversal symmetry were demonstrated using MS on a PhC WG~\cite{Liu2020}. Of particular interest is the potential to realize topological polariton states protected by true time-reversal symmetry breaking~\cite{PhysRevX.5.031001} by combining polarization splitting, photonic crystal bandgap and exciton Zeeman splitting due to external magnetic field. So far similar states have only demonstrated in microcavities in the lasing regime owing to a very small bandgap of 0.1 meV~\cite{Klembt2018}.}

\revision{While modulated bandstructure is one key ingredient for future applications, it is equally important to achieve long polariton lifetimes and propagation distances. In MS and organics the lifetimes are short, which may occur due to inhomogeneous broadening of the exciton linewidth up to $\sim$10 meV. The longest reported polariton lifetimes are achieved in GaAs-based structures~\cite{Sun2017} where, owing to the high quality quantum wells, polariton linewidths can be 10-100 times smaller. Furthermore, a crucial feature of previous works is that the studied states are near zero in-plane momentum where coupling to freely propagating waves in the surroundings inevitable leads to high photonic losses. For example, in Dang et. al.~\cite{Dang2020} even the simulated linewidth for passive materials is of order 16 meV suggesting only a 40 fs photon lifetime. To achieve long polariton lifetime it is critical that states are produced at high wavenumbers where total internal reflection (TIR) prevents radiative loss. For GaAs-based WGs lifetimes (propagation lengths) on the order of 10 ps (500\um{}) can be expected~\cite{Walker2019,doi:10.1063/1.4773590}.
Until now, however, strong bandstructure modulation was not demonstrated in GaAs-based polariton WGs, and polariton PhC states protected by TIR have not so far been demonstrated \revision{in any material system}. Achieving the former is challenging since it relies on etching through a significant fraction of the core. However, etching through or too near the quantum wells leads to high losses associated with surface recombination effects.} While previous works \cite{PhysRevB.80.201308} have demonstrated the persistence of strong coupling in patterned GaAs-based WGs, the modulated region was spatially separated from the WG core leading to only a weak perturbation and no opening of a gap.


\revision{In this work, we implement two-dimensional square lattice PhCs in GaAs-based WGs in the strong coupling regime. We achieve PhC band gaps near the Brillouin zone (BZ) edge where the states are protected by TIR, and show low loss propagation over the $\sim$ 50\um{} wide PhC for states outside the gap. The band gap, with width $\sim$10 meV, \revision{(roughly an order of magnitude larger than in microcavity lattices~\cite{AMO2016934,Schneider_2016})}, is the signature of the polariton bandstructure being strongly modulated by the periodic lattice. Finally, we demonstrate one of the useful features of polariton-based platforms by exploiting the excitonic content of the states to tune the gap width using an external magnetic field.}

\begin{figure}
\centering
\includegraphics[width=0.48\textwidth]{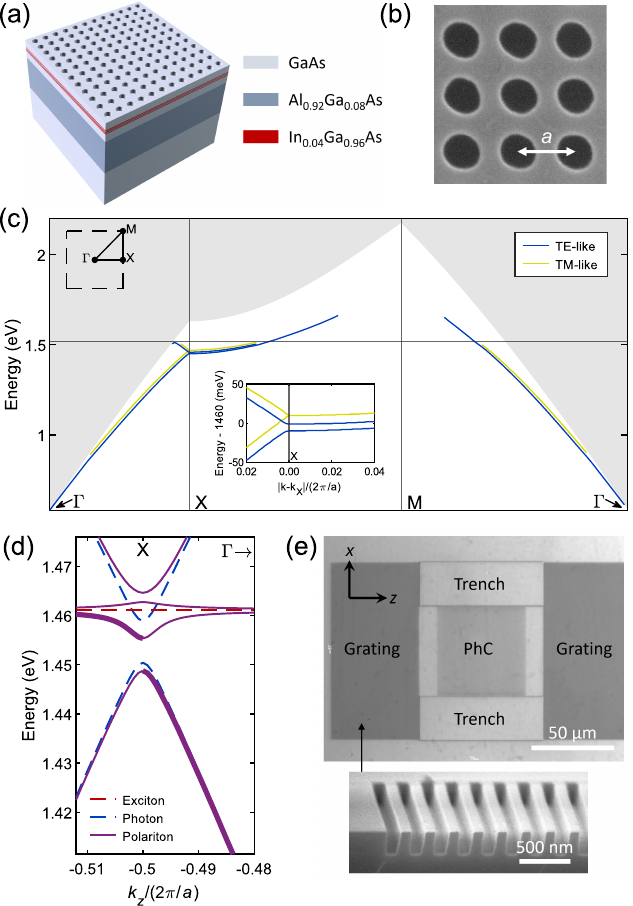}
    \caption{(a) Schematic of photonic crystal (PhC) structure in a slab waveguide. (b) Scanning electron microscope (SEM) image of holes etched into the surface of the waveguide showing the centre-to-centre distance $a$. \revision{(c) Bandstructure of the photon modes of a PhC with $a=126$ nm, $d=47$ nm, $h=64$ nm. The horizontal axis follows the triangular path $\Gamma$-X-M-$\Gamma$ indicated on the inset schematic of the first BZ  (top left). Vertical lines indicate the X and M points. Horizontal line indicates the GaAs bandgap. Central inset is a zoom of the region around the X point.}
    (d) Calculated polariton dispersion relation in the vicinity of the X point ($k_z/(2\pi/a)=-0.5$) for the PhC in (c). The thick lines correspond to the states which are visible in Fig. \ref{fig2}. (e) Optical microscope image of a PhC with grating couplers allowing light to be coupled in and out along the $z$ direction and with etched trenches. The inset shows an angled SEM image of a cleaved etched grating.}
\label{fig1}
\end{figure}

\revision{The sample, illustrated in Fig.~\ref{fig1}(a), is a planar WG into which holes are etched to form a square lattice PhC. The GaAs WG core layer has total thickness 295 nm and contains three 10 nm wide In$_{0.06}$Ga$_{0.94}$As quantum wells (QWs) separated by 10 nm GaAs barriers and placed at the peak of the transverse-electric (TE) polarised field in order to maximize the Rabi coupling. The upper quantum well is 137 nm below the surface, which allows etch depths up to 87 nm ($\sim$30\% of the core thickness) while still leaving 50 nm GaAs cap to protect the wells. The core is separated from the GaAs substrate by a 750 nm thick AlGaAs cladding layer with 90\% average Al composition. An extended discussion of the design of the layer structure is given in the supplementary material.}



\begin{figure*}
\centering
\includegraphics[width=\textwidth]{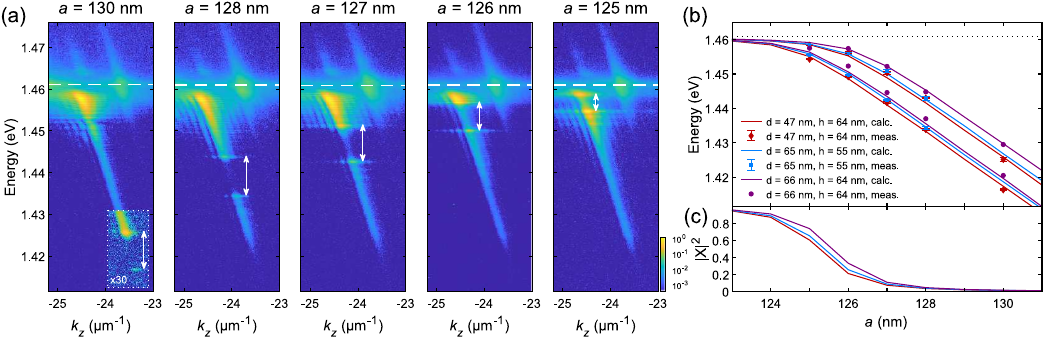}
\caption{(a) Angle-resolved photoluminescence spectra corresponding to PhCs with different periodicities. The devices have holes with diameter d = 47 nm and etch depth h = 64 nm. (b) Measured and calculated positions of gaps (upper and lower band edges) against PhC period for devices with different hole profiles (different diameters d and etch depths h). The calculated positions are given by taking the gaps from photonic simulations and assuming a Rabi splitting of 9 meV with the exciton resonance. The error bars on the measured data points represent the standard deviation from measurements of different devices with the same hole profiles. \revision{(c) Exciton fraction at the center of the gap vs. periodicity for hole profiles corresponding to the solid curves in panel (b).}}
\label{fig2}
\end{figure*}

The samples were grown by MBE and then patterned with a soft mask using electron beam lithography to form PhCs with a square lattice geometry, as shown in Figs. \ref{fig1}(a) and (b). The patterns were etched down into the planar structure using inductively coupled plasma etching with a chlorine/argon chemistry. Several samples were fabricated, with lattice constants $a$ between 120-130 nm, hole etch depths $h$ between 55-64 nm and hole diameters $d$ between 47-66 nm.
\revision{The ranges for fabrication were selected by calculating the PhC bandstructure using Lumerical® FDTD solutions finite difference time domain software (see Supplementary Material for further details). Fig.~\ref{fig1}(c) shows an example of the simulated bandstructure for photons (without coupling to the exciton) across the whole Brillouin zone (BZ). The polarisation of states in slab photonic crystals may be classified as TE-like or TM-like, with electric field mostly in or out of the WG plane respectively~\cite{PhotonicCrystalsBook}. Only the TE-like modes couple strongly to the excitons in our QWs. The bands for both polarisations are shown in Fig.~\ref{fig1}(c). No narrow linewidth states could be identified above the light cone (in the shaded region) or above the GaAs band gap (1.519 eV) since such states are either not guided or strongly absorbed respectively. In Fig.~\ref{fig1}(c) the modes get broader with increasing energy above the band gap and we choose to plot the dispersion of the modes, which have linewidth $<100$ meV. There is a $\sim$ 10 meV gap in the TE-like states around the X point, which we will examine in more detail momentarily. From X towards M the bands increase in frequency so that the band gap does not span all momentum states but exists only for modes propagating near X. A larger gap could be engineered by using a different lattice geometry such as honeycomb~\cite{PhysRevX.5.031001} but even this small gap serves to illustrate that polariton bandstructure can be modulated by many times the linewidth, and could support edge states where the bulk 2D lattice provides other desirable properties such as topological protection~\cite{PhysRevX.5.031001}.}

\revision{The bandstructure was calculated for a range of $a$, $d$ and $h$ and close-to-optimal values were selected for fabrication such that the gap in the TE-like modes at the X point is near the exciton energy and so that the gap width was maximized without etching too close to the QWs. In experiment we will observe polariton rather than pure photon bands. Fig.~\ref{fig1}(d) shows an example of the calculated polariton dispersion resulting from strong coupling between the TE-like mode and QW excitons, focussing on the region around the X point. The polariton band gap can be seen clearly.} 
To enable us to experimentally study gap formation within the PhCs, grating couplers were fabricated on either side of the PhC in the $z$ direction (Fig. \ref{fig1}(c)). These fold the waveguide modes into the radiative region allowing light to be coupled in and out. On the other two sides of the PhC, trenches were etched (through the active region) for prospective studies of edge states, although we note that this is an entirely optional design feature. The side length of the PhCs corresponds to $\sim$400 periods.

For optical measurements, the sample was held at low temperatures (<8 K) and excited by a cw laser at 637 nm focused to a $\sim$3\um{} spot on the PhC surface. This nonresonant excitation incoherently populates the polariton modes of the PhC WGs (which lie below the light cone) by multiple relaxation processes. The polaritons may then propagate out of the PhC to the grating where they are scattered out and recorded using a spectrometer.
Spatial filtering optics were used to collect the emission from selected regions on the sample.  

\begin{figure}
\centering
\includegraphics[width=0.48\textwidth]{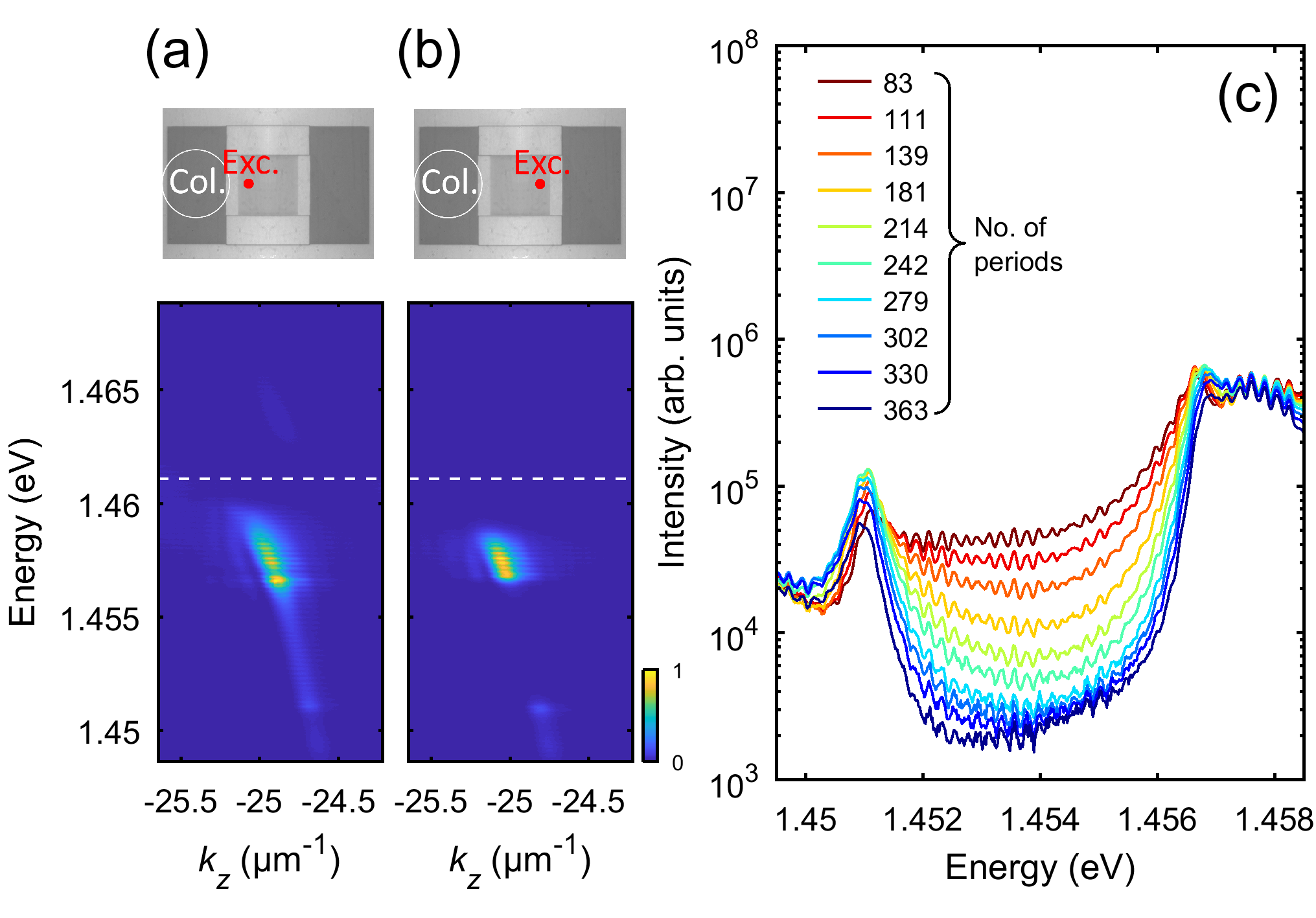}
\caption{Measured angle-resolved PL spectra when the excitation spot (Exc.) is at the near (a) and far (b) edges of the PhC (\textit{a}=126 nm) with respect to the grating from which light is collected (Col.). (c) PL spectra measured for different excitation positions (i.e. number of PhC periods between the excitation and detection spots).}
\label{fig3}
\end{figure}

In Fig. \ref{fig2}(a) we show typical examples of the emission collected from gratings adjacent to the PhCs, measured for a single set of devices and corresponding to a cut through the X point, along the $\Gamma$-X-$\Gamma$ direction in momentum space. The heavy attenuation of PL intensity within particular energy windows directly results from the presence of band gaps in the PhC slabs. Polaritons at energies in the gaps cannot propagate out of the PhC to be detected at the grating. 
We stress that these gaps arise from the periodic potential created by the PhCs, and are qualitatively very different to the resonances induced by back-reflection of guided modes between pairs of gratings in ref.~\onlinecite{Suarez-Forero:20}. 
We see that as the PhC period $a$ is made successively smaller, the gap (shown by the double-sided vertical arrow) moves upwards in energy, becoming smaller as it approaches the exciton resonance (dashed white line). The reduction in the gap size arises from reduction of the photonic fraction as the gap approaches the exciton resonance, confirming that the strong coupling is retained in the PhC region. 
This can also be seen from the anticrossing behaviour of the band gap edges in Fig. \ref{fig2}(b), which also includes results from devices with different hole profiles.
This is in contrast to our simulations of the purely photonic structure (without strong coupling to an exciton) in which the gap size varies little across this range of periods. Since the normal mode (Rabi) splitting between the photonic and excitonic resonances is known, we can calculate the exciton fraction of each gap using 

\begin{equation}
    |X|^2=\frac{1}{2}\left[1+\frac{E_G^{ph}-E_X}{\sqrt{(E_G^{ph}-E_X)^2}+\hbar\Omega^2}\right],
\end{equation}
where $E_G^{ph}$ is the central gap energy of the bare photonic gap, calculated using $E_G^{ph}=\Omega^2/4(E_X-E_G^{pol})+E_G^{pol}$ where $E_G^{pol}$ is the central gap energy of the lower polariton branch. In the data shown in Fig. \ref{fig2}, the gap passes from almost fully photonic ($|X|^2$ = 1\%) to predominantly excitonic ($|X|^2$ = 61\%)\revision{, as can be seen from Fig.~\ref{fig2}(c)}. In other devices we have measured gaps with exciton fractions as high as 73\%. We note that emission from states within $\sim$1.5 meV below the exciton line is heavily attenuated due to absorption which, along with the broadened polariton linewidths, places an upper bound on the exciton fraction of gaps which can be observed in our current samples. This may be improved somewhat by reducing the exciton inhomogeneous linewidth and/or increasing the Rabi splitting to increase $|X|^2$ for polariton states further from the highly absorbing part of the exciton tail. We also note that within the gaps some weak emission is visible, implying that the reflectivity is below unity, which arises from the finite extent of the PhCs. As we show in Fig. \ref{fig3}, the extinction within the gap depends strongly on the length of photonic crystal (i.e. number of lattice periods) between the excitation spot and the edge of the PhC. Increasing the number of periods strongly increases the attenuation for energies in the gap. This is clear evidence that the observed gaps are due to reflection of propagating polaritons from the periodic PhC structure. \revision{We further note that outside the band gap there is little observable dependence on excitation spot distance. This demonstrates that the decay length of freely propagating polaritons in the photonic crystal (but outside the gap) is much larger than the 35\um{} length over which the spot was moved. This is consistent with the $\sim$500\um{} lengths in unpatterned polariton waveguides.~\cite{Walker2019,doi:10.1063/1.4773590}}

\begin{figure}
\centering
\includegraphics[width=0.48\textwidth]{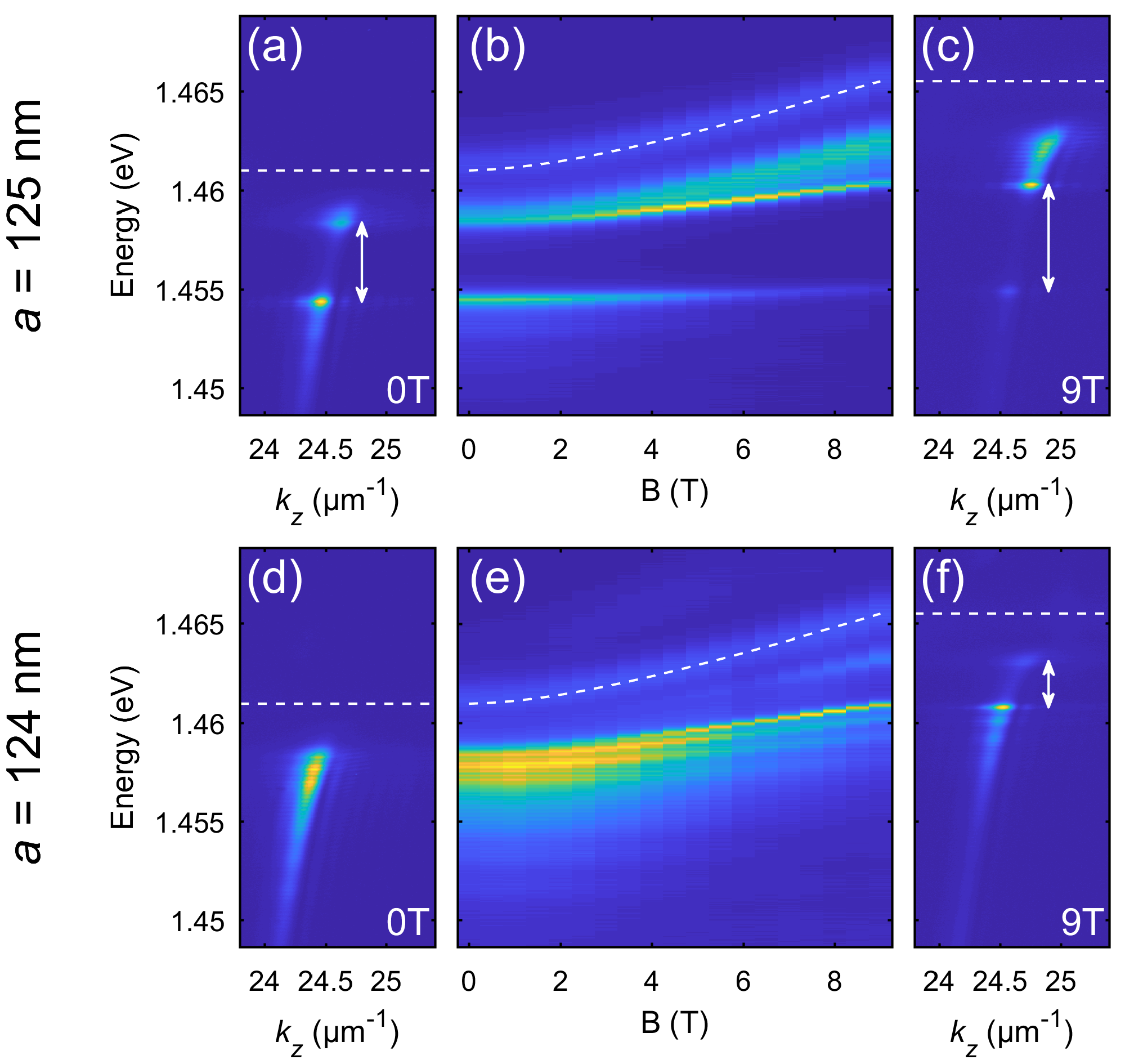}
\caption{(a)-(c) Magnetic field dependence of spectra for PhC with $a$ = 125 nm. The angle-resolved spectra measured at 0 T and 9 T are shown in (a) and (c) respectively, and contour plots of the angle-integrated spectrum are shown in (b). (d)-(f) Same as (a)-(c) for PhC with $a$ = 124 nm. The colour scale is the same as that of Fig. \ref{fig3}.}
\label{fig4}
\end{figure}

In contrast to our system, the primary gaps studied in periodic potentials in Bragg microcavities are typically formed in the photonic part of the spectrum and thus have far smaller exciton fractions, which severely limits the ability to tune the gap position using external fields. If the exciton content is large, however, one may employ diverse methods to tune the energy of the gap including temperature, optical excitation and electric and magnetic fields \cite{Rahimi-Iman2020}. In order to demonstrate the feasibility of tuning gaps in our devices, and also to show unambiguously that they are polaritonic in nature, we placed our sample in a magnetic field in Faraday geometry. The results are summarized in Fig. \ref{fig4}. 
The predominant effect is the diamagnetic (blue)shift of the exciton resonance, which reduces the exciton fraction of the gaps and hence increases their size. For the PhC with $a$ = 125 nm, the gap size can be increased from 4.0 meV at B = 0 T to 5.4 meV at B = +9 T (Figs.\ref{fig4}(a)-(c)). In the case of the PhC with $a$ = 124 nm, the full gap is not visible at B = 0 T, since the upper gap energy lies in the heavily attenuated region below the exciton resonance. At higher magnetic fields however the gap becomes visible, reaching a size of 2.4 meV at B = +9 T (Figs.\ref{fig4}(d)-(f)). \revision{We thus demonstrate that the polariton band structure can be tuned by varying the exciton content using magnetic field. It should be noted that to obtain the larger gap sizes requires reducing the exciton content (see Fig.~\ref{fig2}(c)), which will also reduce the polariton nonlinear interaction strength. The tuning can be used as a lever to strike an optimal balance between these properties.}

For finite magnetic fields there is also a Zeeman splitting of the exciton, which reaches values exceeding 0.5 meV at the highest fields. Thus, the basic ingredients we have presented, namely, gaps in the excitonic part of the spectrum and an exciton Zeeman splitting are those described in ref. \onlinecite{PhysRevX.5.031001} as the criteria for quantum Hall type (chiral) edge states at the boundaries of the system. However, for true topological protection one needs the Zeeman splitting to exceed the size of the gap; this will require an enhancement of the exciton $g$ factor which in our system could be achieved by varying the In composition or width of the quantum wells \cite{PhysRevB.51.7361}. Alternatively one may think about employing semimagnetic quantum wells in a Te-based system \cite{PhysRevB.95.085429} or using transition metal dichalcogenides where spin-dependent strong coupling to a photonic mode can create a giant effective Zeeman splitting exceeding 10 meV \footnote{A. I. Tartakovskii, private communication (unpublished)}. \revision{We note that even without a global band gap the system may support topological edge states with well defined momentum close to the X point. Losses due to scattering into bulk modes can be minimised by maximising the gap size using other lattice geometries such as hexagonal~\cite{PhysRevX.5.031001}}. 

In summary, we have presented a platform to study \revision{strongly modulated} exciton-polariton band structures using patterned slab waveguides in the strong coupling regime. \revision{We observe low loss propagating states outside the light cone and band gaps of order 10 meV.} We have demonstrated that the gaps can be controlled both through the photonic component (varying the period of the crystal) or the excitonic component (external magnetic field). For future studies we envisage the patterning of photonic crystals with different lattice geometries featuring exotic dispersion relations \cite{PhysRevLett.112.116402,PhysRevLett.120.097401}, as well as interfacing waveguides with other excitonic materials such as atomically thin semiconductors \cite{Kravtsov2020} and organic polymers \cite{Jayaprakash2020}. Our system could thus offer a flexible and promising alternative to microcavity-based polariton lattices for the study of topological states and implementation of optoelectronic devices. 



\section*{Supplementary Material}
Further details of the PhC structure and simulations used in the design process are given in the associated supplementary material file.

\begin{acknowledgments}
The work was supported by UK EPSRC Grants EP/N031776/1 and EP/R04385X/1 and by the Russian Science Foundation (Project No. 19-72-20120).
\end{acknowledgments}

\section*{The data availability statement}
The data that support the findings of this study are available from the corresponding author upon reasonable request.

\bibliography{bibliography}

\begin{thebibliography}{33}%
\makeatletter
\providecommand \@ifxundefined [1]{%
 \@ifx{#1\undefined}
}%
\providecommand \@ifnum [1]{%
 \ifnum #1\expandafter \@firstoftwo
 \else \expandafter \@secondoftwo
 \fi
}%
\providecommand \@ifx [1]{%
 \ifx #1\expandafter \@firstoftwo
 \else \expandafter \@secondoftwo
 \fi
}%
\providecommand \natexlab [1]{#1}%
\providecommand \enquote  [1]{``#1''}%
\providecommand \bibnamefont  [1]{#1}%
\providecommand \bibfnamefont [1]{#1}%
\providecommand \citenamefont [1]{#1}%
\providecommand \href@noop [0]{\@secondoftwo}%
\providecommand \href [0]{\begingroup \@sanitize@url \@href}%
\providecommand \@href[1]{\@@startlink{#1}\@@href}%
\providecommand \@@href[1]{\endgroup#1\@@endlink}%
\providecommand \@sanitize@url [0]{\catcode `\\12\catcode `\$12\catcode
  `\&12\catcode `\#12\catcode `\^12\catcode `\_12\catcode `\%12\relax}%
\providecommand \@@startlink[1]{}%
\providecommand \@@endlink[0]{}%
\providecommand \url  [0]{\begingroup\@sanitize@url \@url }%
\providecommand \@url [1]{\endgroup\@href {#1}{\urlprefix }}%
\providecommand \urlprefix  [0]{URL }%
\providecommand \Eprint [0]{\href }%
\providecommand \doibase [0]{http://dx.doi.org/}%
\providecommand \selectlanguage [0]{\@gobble}%
\providecommand \bibinfo  [0]{\@secondoftwo}%
\providecommand \bibfield  [0]{\@secondoftwo}%
\providecommand \translation [1]{[#1]}%
\providecommand \BibitemOpen [0]{}%
\providecommand \bibitemStop [0]{}%
\providecommand \bibitemNoStop [0]{.\EOS\space}%
\providecommand \EOS [0]{\spacefactor3000\relax}%
\providecommand \BibitemShut  [1]{\csname bibitem#1\endcsname}%
\let\auto@bib@innerbib\@empty
\bibitem [{\citenamefont {Sanvitto}\ and\ \citenamefont
  {K{\'e}na-Cohen}(2016)}]{Sanvitto2016}%
  \BibitemOpen
  \bibfield  {author} {\bibinfo {author} {\bibfnamefont {D.}~\bibnamefont
  {Sanvitto}}\ and\ \bibinfo {author} {\bibfnamefont {S.}~\bibnamefont
  {K{\'e}na-Cohen}},\ }\bibfield  {title} {\enquote {\bibinfo {title} {The road
  towards polaritonic devices},}\ }\href {\doibase 10.1038/nmat4668} {\bibfield
   {journal} {\bibinfo  {journal} {Nature Materials}\ }\textbf {\bibinfo
  {volume} {15}},\ \bibinfo {pages} {1061--1073} (\bibinfo {year}
  {2016})}\BibitemShut {NoStop}%
\bibitem [{\citenamefont {Amo}\ and\ \citenamefont {Bloch}(2016)}]{AMO2016934}%
  \BibitemOpen
  \bibfield  {author} {\bibinfo {author} {\bibfnamefont {A.}~\bibnamefont
  {Amo}}\ and\ \bibinfo {author} {\bibfnamefont {J.}~\bibnamefont {Bloch}},\
  }\bibfield  {title} {\enquote {\bibinfo {title} {Exciton-polaritons in
  lattices: A non-linear photonic simulator},}\ }\href {\doibase
  https://doi.org/10.1016/j.crhy.2016.08.007} {\bibfield  {journal} {\bibinfo
  {journal} {Comptes Rendus Physique}\ }\textbf {\bibinfo {volume} {17}},\
  \bibinfo {pages} {934 -- 945} (\bibinfo {year} {2016})},\ \bibinfo {note}
  {polariton physics / Physique des polaritons}\BibitemShut {NoStop}%
\bibitem [{\citenamefont {Schneider}\ \emph {et~al.}(2016)\citenamefont
  {Schneider}, \citenamefont {Winkler}, \citenamefont {Fraser}, \citenamefont
  {Kamp}, \citenamefont {Yamamoto}, \citenamefont {Ostrovskaya},\ and\
  \citenamefont {Höfling}}]{Schneider_2016}%
  \BibitemOpen
  \bibfield  {author} {\bibinfo {author} {\bibfnamefont {C.}~\bibnamefont
  {Schneider}}, \bibinfo {author} {\bibfnamefont {K.}~\bibnamefont {Winkler}},
  \bibinfo {author} {\bibfnamefont {M.~D.}\ \bibnamefont {Fraser}}, \bibinfo
  {author} {\bibfnamefont {M.}~\bibnamefont {Kamp}}, \bibinfo {author}
  {\bibfnamefont {Y.}~\bibnamefont {Yamamoto}}, \bibinfo {author}
  {\bibfnamefont {E.~A.}\ \bibnamefont {Ostrovskaya}}, \ and\ \bibinfo {author}
  {\bibfnamefont {S.}~\bibnamefont {Höfling}},\ }\bibfield  {title} {\enquote
  {\bibinfo {title} {Exciton-polariton trapping and potential landscape
  engineering},}\ }\href {\doibase 10.1088/0034-4885/80/1/016503} {\bibfield
  {journal} {\bibinfo  {journal} {Reports on Progress in Physics}\ }\textbf
  {\bibinfo {volume} {80}},\ \bibinfo {pages} {016503} (\bibinfo {year}
  {2016})}\BibitemShut {NoStop}%
\bibitem [{\citenamefont {Wertz}\ \emph {et~al.}(2012)\citenamefont {Wertz},
  \citenamefont {Amo}, \citenamefont {Solnyshkov}, \citenamefont {Ferrier},
  \citenamefont {Liew}, \citenamefont {Sanvitto}, \citenamefont {Senellart},
  \citenamefont {Sagnes}, \citenamefont {Lema\^{\i}tre}, \citenamefont
  {Kavokin}, \citenamefont {Malpuech},\ and\ \citenamefont
  {Bloch}}]{PhysRevLett.109.216404}%
  \BibitemOpen
  \bibfield  {author} {\bibinfo {author} {\bibfnamefont {E.}~\bibnamefont
  {Wertz}}, \bibinfo {author} {\bibfnamefont {A.}~\bibnamefont {Amo}}, \bibinfo
  {author} {\bibfnamefont {D.~D.}\ \bibnamefont {Solnyshkov}}, \bibinfo
  {author} {\bibfnamefont {L.}~\bibnamefont {Ferrier}}, \bibinfo {author}
  {\bibfnamefont {T.~C.~H.}\ \bibnamefont {Liew}}, \bibinfo {author}
  {\bibfnamefont {D.}~\bibnamefont {Sanvitto}}, \bibinfo {author}
  {\bibfnamefont {P.}~\bibnamefont {Senellart}}, \bibinfo {author}
  {\bibfnamefont {I.}~\bibnamefont {Sagnes}}, \bibinfo {author} {\bibfnamefont
  {A.}~\bibnamefont {Lema\^{\i}tre}}, \bibinfo {author} {\bibfnamefont {A.~V.}\
  \bibnamefont {Kavokin}}, \bibinfo {author} {\bibfnamefont {G.}~\bibnamefont
  {Malpuech}}, \ and\ \bibinfo {author} {\bibfnamefont {J.}~\bibnamefont
  {Bloch}},\ }\bibfield  {title} {\enquote {\bibinfo {title} {Propagation and
  amplification dynamics of 1d polariton condensates},}\ }\href {\doibase
  10.1103/PhysRevLett.109.216404} {\bibfield  {journal} {\bibinfo  {journal}
  {Phys. Rev. Lett.}\ }\textbf {\bibinfo {volume} {109}},\ \bibinfo {pages}
  {216404} (\bibinfo {year} {2012})}\BibitemShut {NoStop}%
\bibitem [{\citenamefont {Nguyen}\ \emph {et~al.}(2013)\citenamefont {Nguyen},
  \citenamefont {Vishnevsky}, \citenamefont {Sturm}, \citenamefont {Tanese},
  \citenamefont {Solnyshkov}, \citenamefont {Galopin}, \citenamefont
  {Lema\^{\i}tre}, \citenamefont {Sagnes}, \citenamefont {Amo}, \citenamefont
  {Malpuech},\ and\ \citenamefont {Bloch}}]{PhysRevLett.110.236601}%
  \BibitemOpen
  \bibfield  {author} {\bibinfo {author} {\bibfnamefont {H.~S.}\ \bibnamefont
  {Nguyen}}, \bibinfo {author} {\bibfnamefont {D.}~\bibnamefont {Vishnevsky}},
  \bibinfo {author} {\bibfnamefont {C.}~\bibnamefont {Sturm}}, \bibinfo
  {author} {\bibfnamefont {D.}~\bibnamefont {Tanese}}, \bibinfo {author}
  {\bibfnamefont {D.}~\bibnamefont {Solnyshkov}}, \bibinfo {author}
  {\bibfnamefont {E.}~\bibnamefont {Galopin}}, \bibinfo {author} {\bibfnamefont
  {A.}~\bibnamefont {Lema\^{\i}tre}}, \bibinfo {author} {\bibfnamefont
  {I.}~\bibnamefont {Sagnes}}, \bibinfo {author} {\bibfnamefont
  {A.}~\bibnamefont {Amo}}, \bibinfo {author} {\bibfnamefont {G.}~\bibnamefont
  {Malpuech}}, \ and\ \bibinfo {author} {\bibfnamefont {J.}~\bibnamefont
  {Bloch}},\ }\bibfield  {title} {\enquote {\bibinfo {title} {Realization of a
  double-barrier resonant tunneling diode for cavity polaritons},}\ }\href
  {\doibase 10.1103/PhysRevLett.110.236601} {\bibfield  {journal} {\bibinfo
  {journal} {Phys. Rev. Lett.}\ }\textbf {\bibinfo {volume} {110}},\ \bibinfo
  {pages} {236601} (\bibinfo {year} {2013})}\BibitemShut {NoStop}%
\bibitem [{\citenamefont {Sturm}\ \emph {et~al.}(2014)\citenamefont {Sturm},
  \citenamefont {Tanese}, \citenamefont {Nguyen}, \citenamefont {Flayac},
  \citenamefont {Galopin}, \citenamefont {Lema{\^i}tre}, \citenamefont
  {Sagnes}, \citenamefont {Solnyshkov}, \citenamefont {Amo}, \citenamefont
  {Malpuech},\ and\ \citenamefont {Bloch}}]{Sturm2014}%
  \BibitemOpen
  \bibfield  {author} {\bibinfo {author} {\bibfnamefont {C.}~\bibnamefont
  {Sturm}}, \bibinfo {author} {\bibfnamefont {D.}~\bibnamefont {Tanese}},
  \bibinfo {author} {\bibfnamefont {H.~S.}\ \bibnamefont {Nguyen}}, \bibinfo
  {author} {\bibfnamefont {H.}~\bibnamefont {Flayac}}, \bibinfo {author}
  {\bibfnamefont {E.}~\bibnamefont {Galopin}}, \bibinfo {author} {\bibfnamefont
  {A.}~\bibnamefont {Lema{\^i}tre}}, \bibinfo {author} {\bibfnamefont
  {I.}~\bibnamefont {Sagnes}}, \bibinfo {author} {\bibfnamefont
  {D.}~\bibnamefont {Solnyshkov}}, \bibinfo {author} {\bibfnamefont
  {A.}~\bibnamefont {Amo}}, \bibinfo {author} {\bibfnamefont {G.}~\bibnamefont
  {Malpuech}}, \ and\ \bibinfo {author} {\bibfnamefont {J.}~\bibnamefont
  {Bloch}},\ }\bibfield  {title} {\enquote {\bibinfo {title} {All-optical phase
  modulation in a cavity-polariton mach--zehnder interferometer},}\ }\href
  {\doibase 10.1038/ncomms4278} {\bibfield  {journal} {\bibinfo  {journal}
  {Nature Communications}\ }\textbf {\bibinfo {volume} {5}},\ \bibinfo {pages}
  {3278} (\bibinfo {year} {2014})}\BibitemShut {NoStop}%
\bibitem [{\citenamefont {Marsault}\ \emph {et~al.}(2015)\citenamefont
  {Marsault}, \citenamefont {Nguyen}, \citenamefont {Tanese}, \citenamefont
  {Lemaître}, \citenamefont {Galopin}, \citenamefont {Sagnes}, \citenamefont
  {Amo},\ and\ \citenamefont {Bloch}}]{doi:10.1063/1.4936158}%
  \BibitemOpen
  \bibfield  {author} {\bibinfo {author} {\bibfnamefont {F.}~\bibnamefont
  {Marsault}}, \bibinfo {author} {\bibfnamefont {H.~S.}\ \bibnamefont
  {Nguyen}}, \bibinfo {author} {\bibfnamefont {D.}~\bibnamefont {Tanese}},
  \bibinfo {author} {\bibfnamefont {A.}~\bibnamefont {Lemaître}}, \bibinfo
  {author} {\bibfnamefont {E.}~\bibnamefont {Galopin}}, \bibinfo {author}
  {\bibfnamefont {I.}~\bibnamefont {Sagnes}}, \bibinfo {author} {\bibfnamefont
  {A.}~\bibnamefont {Amo}}, \ and\ \bibinfo {author} {\bibfnamefont
  {J.}~\bibnamefont {Bloch}},\ }\bibfield  {title} {\enquote {\bibinfo {title}
  {Realization of an all optical exciton-polariton router},}\ }\href {\doibase
  10.1063/1.4936158} {\bibfield  {journal} {\bibinfo  {journal} {Applied
  Physics Letters}\ }\textbf {\bibinfo {volume} {107}},\ \bibinfo {pages}
  {201115} (\bibinfo {year} {2015})},\ \Eprint
  {http://arxiv.org/abs/https://doi.org/10.1063/1.4936158}
  {https://doi.org/10.1063/1.4936158} \BibitemShut {NoStop}%
\bibitem [{\citenamefont {St-Jean}\ \emph {et~al.}(2017)\citenamefont
  {St-Jean}, \citenamefont {Goblot}, \citenamefont {Galopin}, \citenamefont
  {Lema{\^i}tre}, \citenamefont {Ozawa}, \citenamefont {Le~Gratiet},
  \citenamefont {Sagnes}, \citenamefont {Bloch},\ and\ \citenamefont
  {Amo}}]{St-Jean2017}%
  \BibitemOpen
  \bibfield  {author} {\bibinfo {author} {\bibfnamefont {P.}~\bibnamefont
  {St-Jean}}, \bibinfo {author} {\bibfnamefont {V.}~\bibnamefont {Goblot}},
  \bibinfo {author} {\bibfnamefont {E.}~\bibnamefont {Galopin}}, \bibinfo
  {author} {\bibfnamefont {A.}~\bibnamefont {Lema{\^i}tre}}, \bibinfo {author}
  {\bibfnamefont {T.}~\bibnamefont {Ozawa}}, \bibinfo {author} {\bibfnamefont
  {L.}~\bibnamefont {Le~Gratiet}}, \bibinfo {author} {\bibfnamefont
  {I.}~\bibnamefont {Sagnes}}, \bibinfo {author} {\bibfnamefont
  {J.}~\bibnamefont {Bloch}}, \ and\ \bibinfo {author} {\bibfnamefont
  {A.}~\bibnamefont {Amo}},\ }\bibfield  {title} {\enquote {\bibinfo {title}
  {Lasing in topological edge states of a one-dimensional lattice},}\ }\href
  {\doibase 10.1038/s41566-017-0006-2} {\bibfield  {journal} {\bibinfo
  {journal} {Nature Photonics}\ }\textbf {\bibinfo {volume} {11}},\ \bibinfo
  {pages} {651--656} (\bibinfo {year} {2017})}\BibitemShut {NoStop}%
\bibitem [{\citenamefont {Klembt}\ \emph {et~al.}(2018)\citenamefont {Klembt},
  \citenamefont {Harder}, \citenamefont {Egorov}, \citenamefont {Winkler},
  \citenamefont {Ge}, \citenamefont {Bandres}, \citenamefont {Emmerling},
  \citenamefont {Worschech}, \citenamefont {Liew}, \citenamefont {Segev},
  \citenamefont {Schneider},\ and\ \citenamefont {H{\"o}fling}}]{Klembt2018}%
  \BibitemOpen
  \bibfield  {author} {\bibinfo {author} {\bibfnamefont {S.}~\bibnamefont
  {Klembt}}, \bibinfo {author} {\bibfnamefont {T.~H.}\ \bibnamefont {Harder}},
  \bibinfo {author} {\bibfnamefont {O.~A.}\ \bibnamefont {Egorov}}, \bibinfo
  {author} {\bibfnamefont {K.}~\bibnamefont {Winkler}}, \bibinfo {author}
  {\bibfnamefont {R.}~\bibnamefont {Ge}}, \bibinfo {author} {\bibfnamefont
  {M.~A.}\ \bibnamefont {Bandres}}, \bibinfo {author} {\bibfnamefont
  {M.}~\bibnamefont {Emmerling}}, \bibinfo {author} {\bibfnamefont
  {L.}~\bibnamefont {Worschech}}, \bibinfo {author} {\bibfnamefont {T.~C.~H.}\
  \bibnamefont {Liew}}, \bibinfo {author} {\bibfnamefont {M.}~\bibnamefont
  {Segev}}, \bibinfo {author} {\bibfnamefont {C.}~\bibnamefont {Schneider}}, \
  and\ \bibinfo {author} {\bibfnamefont {S.}~\bibnamefont {H{\"o}fling}},\
  }\bibfield  {title} {\enquote {\bibinfo {title} {Exciton-polariton
  topological insulator},}\ }\href {\doibase 10.1038/s41586-018-0601-5}
  {\bibfield  {journal} {\bibinfo  {journal} {Nature}\ }\textbf {\bibinfo
  {volume} {562}},\ \bibinfo {pages} {552--556} (\bibinfo {year}
  {2018})}\BibitemShut {NoStop}%
\bibitem [{\citenamefont {Walker}\ \emph {et~al.}(2013)\citenamefont {Walker},
  \citenamefont {Tinkler}, \citenamefont {Durska}, \citenamefont {Whittaker},
  \citenamefont {Luxmoore}, \citenamefont {Royall}, \citenamefont
  {Krizhanovskii}, \citenamefont {Skolnick}, \citenamefont {Farrer},\ and\
  \citenamefont {Ritchie}}]{doi:10.1063/1.4773590}%
  \BibitemOpen
  \bibfield  {author} {\bibinfo {author} {\bibfnamefont {P.~M.}\ \bibnamefont
  {Walker}}, \bibinfo {author} {\bibfnamefont {L.}~\bibnamefont {Tinkler}},
  \bibinfo {author} {\bibfnamefont {M.}~\bibnamefont {Durska}}, \bibinfo
  {author} {\bibfnamefont {D.~M.}\ \bibnamefont {Whittaker}}, \bibinfo {author}
  {\bibfnamefont {I.~J.}\ \bibnamefont {Luxmoore}}, \bibinfo {author}
  {\bibfnamefont {B.}~\bibnamefont {Royall}}, \bibinfo {author} {\bibfnamefont
  {D.~N.}\ \bibnamefont {Krizhanovskii}}, \bibinfo {author} {\bibfnamefont
  {M.~S.}\ \bibnamefont {Skolnick}}, \bibinfo {author} {\bibfnamefont
  {I.}~\bibnamefont {Farrer}}, \ and\ \bibinfo {author} {\bibfnamefont {D.~A.}\
  \bibnamefont {Ritchie}},\ }\bibfield  {title} {\enquote {\bibinfo {title}
  {Exciton polaritons in semiconductor waveguides},}\ }\href {\doibase
  10.1063/1.4773590} {\bibfield  {journal} {\bibinfo  {journal} {Applied
  Physics Letters}\ }\textbf {\bibinfo {volume} {102}},\ \bibinfo {pages}
  {012109} (\bibinfo {year} {2013})},\ \Eprint
  {http://arxiv.org/abs/https://doi.org/10.1063/1.4773590}
  {https://doi.org/10.1063/1.4773590} \BibitemShut {NoStop}%
\bibitem [{\citenamefont {Jamadi}\ \emph {et~al.}(2018)\citenamefont {Jamadi},
  \citenamefont {Reveret}, \citenamefont {Disseix}, \citenamefont {Medard},
  \citenamefont {Leymarie}, \citenamefont {Moreau}, \citenamefont {Solnyshkov},
  \citenamefont {Deparis}, \citenamefont {Leroux}, \citenamefont {Cambril},
  \citenamefont {Bouchoule}, \citenamefont {Zuniga-Perez},\ and\ \citenamefont
  {Malpuech}}]{Jamadi2018}%
  \BibitemOpen
  \bibfield  {author} {\bibinfo {author} {\bibfnamefont {O.}~\bibnamefont
  {Jamadi}}, \bibinfo {author} {\bibfnamefont {F.}~\bibnamefont {Reveret}},
  \bibinfo {author} {\bibfnamefont {P.}~\bibnamefont {Disseix}}, \bibinfo
  {author} {\bibfnamefont {F.}~\bibnamefont {Medard}}, \bibinfo {author}
  {\bibfnamefont {J.}~\bibnamefont {Leymarie}}, \bibinfo {author}
  {\bibfnamefont {A.}~\bibnamefont {Moreau}}, \bibinfo {author} {\bibfnamefont
  {D.}~\bibnamefont {Solnyshkov}}, \bibinfo {author} {\bibfnamefont
  {C.}~\bibnamefont {Deparis}}, \bibinfo {author} {\bibfnamefont
  {M.}~\bibnamefont {Leroux}}, \bibinfo {author} {\bibfnamefont
  {E.}~\bibnamefont {Cambril}}, \bibinfo {author} {\bibfnamefont
  {S.}~\bibnamefont {Bouchoule}}, \bibinfo {author} {\bibfnamefont
  {J.}~\bibnamefont {Zuniga-Perez}}, \ and\ \bibinfo {author} {\bibfnamefont
  {G.}~\bibnamefont {Malpuech}},\ }\bibfield  {title} {\enquote {\bibinfo
  {title} {Edge-emitting polariton laser and amplifier based on a zno
  waveguide},}\ }\href {\doibase 10.1038/s41377-018-0084-z} {\bibfield
  {journal} {\bibinfo  {journal} {Light: Science {\&} Applications}\ }\textbf
  {\bibinfo {volume} {7}},\ \bibinfo {pages} {82} (\bibinfo {year}
  {2018})}\BibitemShut {NoStop}%
\bibitem [{\citenamefont {Su\'{a}rez-Forero}\ \emph {et~al.}(2020)\citenamefont
  {Su\'{a}rez-Forero}, \citenamefont {Riminucci}, \citenamefont {Ardizzone},
  \citenamefont {Giorgi}, \citenamefont {Dominici}, \citenamefont {Todisco},
  \citenamefont {Lerario}, \citenamefont {Pfeiffer}, \citenamefont {Gigli},
  \citenamefont {Ballarini},\ and\ \citenamefont
  {Sanvitto}}]{Suarez-Forero:20}%
  \BibitemOpen
  \bibfield  {author} {\bibinfo {author} {\bibfnamefont {D.~G.}\ \bibnamefont
  {Su\'{a}rez-Forero}}, \bibinfo {author} {\bibfnamefont {F.}~\bibnamefont
  {Riminucci}}, \bibinfo {author} {\bibfnamefont {V.}~\bibnamefont
  {Ardizzone}}, \bibinfo {author} {\bibfnamefont {M.~D.}\ \bibnamefont
  {Giorgi}}, \bibinfo {author} {\bibfnamefont {L.}~\bibnamefont {Dominici}},
  \bibinfo {author} {\bibfnamefont {F.}~\bibnamefont {Todisco}}, \bibinfo
  {author} {\bibfnamefont {G.}~\bibnamefont {Lerario}}, \bibinfo {author}
  {\bibfnamefont {L.~N.}\ \bibnamefont {Pfeiffer}}, \bibinfo {author}
  {\bibfnamefont {G.}~\bibnamefont {Gigli}}, \bibinfo {author} {\bibfnamefont
  {D.}~\bibnamefont {Ballarini}}, \ and\ \bibinfo {author} {\bibfnamefont
  {D.}~\bibnamefont {Sanvitto}},\ }\bibfield  {title} {\enquote {\bibinfo
  {title} {Electrically controlled waveguide polariton laser},}\ }\href
  {\doibase 10.1364/OPTICA.403558} {\bibfield  {journal} {\bibinfo  {journal}
  {Optica}\ }\textbf {\bibinfo {volume} {7}},\ \bibinfo {pages} {1579--1586}
  (\bibinfo {year} {2020})}\BibitemShut {NoStop}%
\bibitem [{\citenamefont {Walker}\ \emph {et~al.}(2019)\citenamefont {Walker},
  \citenamefont {Whittaker}, \citenamefont {Skryabin}, \citenamefont
  {Cancellieri}, \citenamefont {Royall}, \citenamefont {Sich}, \citenamefont
  {Farrer}, \citenamefont {Ritchie}, \citenamefont {Skolnick},\ and\
  \citenamefont {Krizhanovskii}}]{Walker2019}%
  \BibitemOpen
  \bibfield  {author} {\bibinfo {author} {\bibfnamefont {P.~M.}\ \bibnamefont
  {Walker}}, \bibinfo {author} {\bibfnamefont {C.~E.}\ \bibnamefont
  {Whittaker}}, \bibinfo {author} {\bibfnamefont {D.~V.}\ \bibnamefont
  {Skryabin}}, \bibinfo {author} {\bibfnamefont {E.}~\bibnamefont
  {Cancellieri}}, \bibinfo {author} {\bibfnamefont {B.}~\bibnamefont {Royall}},
  \bibinfo {author} {\bibfnamefont {M.}~\bibnamefont {Sich}}, \bibinfo {author}
  {\bibfnamefont {I.}~\bibnamefont {Farrer}}, \bibinfo {author} {\bibfnamefont
  {D.~A.}\ \bibnamefont {Ritchie}}, \bibinfo {author} {\bibfnamefont {M.~S.}\
  \bibnamefont {Skolnick}}, \ and\ \bibinfo {author} {\bibfnamefont {D.~N.}\
  \bibnamefont {Krizhanovskii}},\ }\bibfield  {title} {\enquote {\bibinfo
  {title} {Spatiotemporal continuum generation in polariton waveguides},}\
  }\href {\doibase 10.1038/s41377-019-0120-7} {\bibfield  {journal} {\bibinfo
  {journal} {Light: Science {\&} Applications}\ }\textbf {\bibinfo {volume}
  {8}},\ \bibinfo {pages} {6} (\bibinfo {year} {2019})}\BibitemShut {NoStop}%
\bibitem [{\citenamefont {Paola}\ \emph {et~al.}(2020)\citenamefont {Paola},
  \citenamefont {Walker}, \citenamefont {Emmanuele}, \citenamefont {Yulin},
  \citenamefont {Ciers}, \citenamefont {Zaidi}, \citenamefont {Carlin},
  \citenamefont {Grandjean}, \citenamefont {Shelykh}, \citenamefont {Skolnick},
  \citenamefont {Butté},\ and\ \citenamefont {Krizhanovskii}}]{2009.02059}%
  \BibitemOpen
  \bibfield  {author} {\bibinfo {author} {\bibfnamefont {D.~M.~D.}\
  \bibnamefont {Paola}}, \bibinfo {author} {\bibfnamefont {P.~M.}\ \bibnamefont
  {Walker}}, \bibinfo {author} {\bibfnamefont {R.~P.~A.}\ \bibnamefont
  {Emmanuele}}, \bibinfo {author} {\bibfnamefont {A.~V.}\ \bibnamefont
  {Yulin}}, \bibinfo {author} {\bibfnamefont {J.}~\bibnamefont {Ciers}},
  \bibinfo {author} {\bibfnamefont {Z.}~\bibnamefont {Zaidi}}, \bibinfo
  {author} {\bibfnamefont {J.-F.}\ \bibnamefont {Carlin}}, \bibinfo {author}
  {\bibfnamefont {N.}~\bibnamefont {Grandjean}}, \bibinfo {author}
  {\bibfnamefont {I.}~\bibnamefont {Shelykh}}, \bibinfo {author} {\bibfnamefont
  {M.~S.}\ \bibnamefont {Skolnick}}, \bibinfo {author} {\bibfnamefont
  {R.}~\bibnamefont {Butté}}, \ and\ \bibinfo {author} {\bibfnamefont {D.~N.}\
  \bibnamefont {Krizhanovskii}},\ }\href@noop {} {\enquote {\bibinfo {title}
  {Ultrafast-nonlinear ultraviolet pulse modulation in an alingan polariton
  waveguide operating up to room temperature},}\ } (\bibinfo {year} {2020}),\
  \Eprint {http://arxiv.org/abs/arXiv:2009.02059} {arXiv:2009.02059}
  \BibitemShut {NoStop}%
\bibitem [{\citenamefont {Rosenberg}\ \emph {et~al.}(2018)\citenamefont
  {Rosenberg}, \citenamefont {Liran}, \citenamefont {Mazuz-Harpaz},
  \citenamefont {West}, \citenamefont {Pfeiffer},\ and\ \citenamefont
  {Rapaport}}]{Rosenberg2018}%
  \BibitemOpen
  \bibfield  {author} {\bibinfo {author} {\bibfnamefont {I.}~\bibnamefont
  {Rosenberg}}, \bibinfo {author} {\bibfnamefont {D.}~\bibnamefont {Liran}},
  \bibinfo {author} {\bibfnamefont {Y.}~\bibnamefont {Mazuz-Harpaz}}, \bibinfo
  {author} {\bibfnamefont {K.}~\bibnamefont {West}}, \bibinfo {author}
  {\bibfnamefont {L.}~\bibnamefont {Pfeiffer}}, \ and\ \bibinfo {author}
  {\bibfnamefont {R.}~\bibnamefont {Rapaport}},\ }\bibfield  {title} {\enquote
  {\bibinfo {title} {Strongly interacting dipolar-polaritons},}\ }\href
  {\doibase 10.1126/sciadv.aat8880} {\bibfield  {journal} {\bibinfo  {journal}
  {Science Advances}\ }\textbf {\bibinfo {volume} {4}},\ \bibinfo {pages}
  {eaat8880} (\bibinfo {year} {2018})}\BibitemShut {NoStop}%
\bibitem [{\citenamefont {Su{\'{a}}rez-Forero}\ \emph
  {et~al.}(2021)\citenamefont {Su{\'{a}}rez-Forero}, \citenamefont {Riminucci},
  \citenamefont {Ardizzone}, \citenamefont {Karpowicz}, \citenamefont
  {Maggiolini}, \citenamefont {Macorini}, \citenamefont {Lerario},
  \citenamefont {Todisco}, \citenamefont {Giorgi}, \citenamefont {Dominici},
  \citenamefont {Ballarini}, \citenamefont {Gigli}, \citenamefont {Lanotte},
  \citenamefont {West}, \citenamefont {Baldwin}, \citenamefont {Pfeiffer},\
  and\ \citenamefont {Sanvitto}}]{SuarezForero2021}%
  \BibitemOpen
  \bibfield  {author} {\bibinfo {author} {\bibfnamefont {D.}~\bibnamefont
  {Su{\'{a}}rez-Forero}}, \bibinfo {author} {\bibfnamefont {F.}~\bibnamefont
  {Riminucci}}, \bibinfo {author} {\bibfnamefont {V.}~\bibnamefont
  {Ardizzone}}, \bibinfo {author} {\bibfnamefont {N.}~\bibnamefont
  {Karpowicz}}, \bibinfo {author} {\bibfnamefont {E.}~\bibnamefont
  {Maggiolini}}, \bibinfo {author} {\bibfnamefont {G.}~\bibnamefont
  {Macorini}}, \bibinfo {author} {\bibfnamefont {G.}~\bibnamefont {Lerario}},
  \bibinfo {author} {\bibfnamefont {F.}~\bibnamefont {Todisco}}, \bibinfo
  {author} {\bibfnamefont {M.~D.}\ \bibnamefont {Giorgi}}, \bibinfo {author}
  {\bibfnamefont {L.}~\bibnamefont {Dominici}}, \bibinfo {author}
  {\bibfnamefont {D.}~\bibnamefont {Ballarini}}, \bibinfo {author}
  {\bibfnamefont {G.}~\bibnamefont {Gigli}}, \bibinfo {author} {\bibfnamefont
  {A.}~\bibnamefont {Lanotte}}, \bibinfo {author} {\bibfnamefont
  {K.}~\bibnamefont {West}}, \bibinfo {author} {\bibfnamefont {K.}~\bibnamefont
  {Baldwin}}, \bibinfo {author} {\bibfnamefont {L.}~\bibnamefont {Pfeiffer}}, \
  and\ \bibinfo {author} {\bibfnamefont {D.}~\bibnamefont {Sanvitto}},\
  }\bibfield  {title} {\enquote {\bibinfo {title} {Enhancement of parametric
  effects in polariton waveguides induced by dipolar interactions},}\ }\href
  {\doibase 10.1103/physrevlett.126.137401} {\bibfield  {journal} {\bibinfo
  {journal} {Physical Review Letters}\ }\textbf {\bibinfo {volume} {126}},\
  \bibinfo {pages} {137401} (\bibinfo {year} {2021})}\BibitemShut {NoStop}%
\bibitem [{\citenamefont {Bajoni}\ \emph {et~al.}(2009)\citenamefont {Bajoni},
  \citenamefont {Gerace}, \citenamefont {Galli}, \citenamefont {Bloch},
  \citenamefont {Braive}, \citenamefont {Sagnes}, \citenamefont {Miard},
  \citenamefont {Lema\^{\i}tre}, \citenamefont {Patrini},\ and\ \citenamefont
  {Andreani}}]{PhysRevB.80.201308}%
  \BibitemOpen
  \bibfield  {author} {\bibinfo {author} {\bibfnamefont {D.}~\bibnamefont
  {Bajoni}}, \bibinfo {author} {\bibfnamefont {D.}~\bibnamefont {Gerace}},
  \bibinfo {author} {\bibfnamefont {M.}~\bibnamefont {Galli}}, \bibinfo
  {author} {\bibfnamefont {J.}~\bibnamefont {Bloch}}, \bibinfo {author}
  {\bibfnamefont {R.}~\bibnamefont {Braive}}, \bibinfo {author} {\bibfnamefont
  {I.}~\bibnamefont {Sagnes}}, \bibinfo {author} {\bibfnamefont
  {A.}~\bibnamefont {Miard}}, \bibinfo {author} {\bibfnamefont
  {A.}~\bibnamefont {Lema\^{\i}tre}}, \bibinfo {author} {\bibfnamefont
  {M.}~\bibnamefont {Patrini}}, \ and\ \bibinfo {author} {\bibfnamefont
  {L.~C.}\ \bibnamefont {Andreani}},\ }\bibfield  {title} {\enquote {\bibinfo
  {title} {Exciton polaritons in two-dimensional photonic crystals},}\ }\href
  {\doibase 10.1103/PhysRevB.80.201308} {\bibfield  {journal} {\bibinfo
  {journal} {Phys. Rev. B}\ }\textbf {\bibinfo {volume} {80}},\ \bibinfo
  {pages} {201308} (\bibinfo {year} {2009})}\BibitemShut {NoStop}%
\bibitem [{\citenamefont {Gerace}\ and\ \citenamefont
  {Andreani}(2007)}]{Gerace2007}%
  \BibitemOpen
  \bibfield  {author} {\bibinfo {author} {\bibfnamefont {D.}~\bibnamefont
  {Gerace}}\ and\ \bibinfo {author} {\bibfnamefont {L.~C.}\ \bibnamefont
  {Andreani}},\ }\bibfield  {title} {\enquote {\bibinfo {title} {Quantum theory
  of exciton-photon coupling in photonic crystal slabs with embedded quantum
  wells},}\ }\href {\doibase 10.1103/physrevb.75.235325} {\bibfield  {journal}
  {\bibinfo  {journal} {Physical Review B}\ }\textbf {\bibinfo {volume} {75}},\
  \bibinfo {pages} {235325} (\bibinfo {year} {2007})}\BibitemShut {NoStop}%
\bibitem [{\citenamefont {Zhang}\ \emph {et~al.}(2018)\citenamefont {Zhang},
  \citenamefont {Gogna}, \citenamefont {Burg}, \citenamefont {Tutuc},\ and\
  \citenamefont {Deng}}]{Zhang2018}%
  \BibitemOpen
  \bibfield  {author} {\bibinfo {author} {\bibfnamefont {L.}~\bibnamefont
  {Zhang}}, \bibinfo {author} {\bibfnamefont {R.}~\bibnamefont {Gogna}},
  \bibinfo {author} {\bibfnamefont {W.}~\bibnamefont {Burg}}, \bibinfo {author}
  {\bibfnamefont {E.}~\bibnamefont {Tutuc}}, \ and\ \bibinfo {author}
  {\bibfnamefont {H.}~\bibnamefont {Deng}},\ }\bibfield  {title} {\enquote
  {\bibinfo {title} {Photonic-crystal exciton-polaritons in monolayer
  semiconductors},}\ }\href {\doibase 10.1038/s41467-018-03188-x} {\bibfield
  {journal} {\bibinfo  {journal} {Nature Communications}\ }\textbf {\bibinfo
  {volume} {9}} (\bibinfo {year} {2018}),\
  10.1038/s41467-018-03188-x}\BibitemShut {NoStop}%
\bibitem [{\citenamefont {Chen}\ \emph {et~al.}(2020)\citenamefont {Chen},
  \citenamefont {Miao}, \citenamefont {Wang}, \citenamefont {Zhong},
  \citenamefont {Saxena}, \citenamefont {Chow}, \citenamefont {Whitehead},
  \citenamefont {Gerace}, \citenamefont {Xu}, \citenamefont {Shi},\ and\
  \citenamefont {Majumdar}}]{Chen2020}%
  \BibitemOpen
  \bibfield  {author} {\bibinfo {author} {\bibfnamefont {Y.}~\bibnamefont
  {Chen}}, \bibinfo {author} {\bibfnamefont {S.}~\bibnamefont {Miao}}, \bibinfo
  {author} {\bibfnamefont {T.}~\bibnamefont {Wang}}, \bibinfo {author}
  {\bibfnamefont {D.}~\bibnamefont {Zhong}}, \bibinfo {author} {\bibfnamefont
  {A.}~\bibnamefont {Saxena}}, \bibinfo {author} {\bibfnamefont
  {C.}~\bibnamefont {Chow}}, \bibinfo {author} {\bibfnamefont {J.}~\bibnamefont
  {Whitehead}}, \bibinfo {author} {\bibfnamefont {D.}~\bibnamefont {Gerace}},
  \bibinfo {author} {\bibfnamefont {X.}~\bibnamefont {Xu}}, \bibinfo {author}
  {\bibfnamefont {S.-F.}\ \bibnamefont {Shi}}, \ and\ \bibinfo {author}
  {\bibfnamefont {A.}~\bibnamefont {Majumdar}},\ }\bibfield  {title} {\enquote
  {\bibinfo {title} {Metasurface integrated monolayer exciton polariton},}\
  }\href {\doibase 10.1021/acs.nanolett.0c01624} {\bibfield  {journal}
  {\bibinfo  {journal} {Nano Letters}\ }\textbf {\bibinfo {volume} {20}},\
  \bibinfo {pages} {5292--5300} (\bibinfo {year} {2020})}\BibitemShut {NoStop}%
\bibitem [{\citenamefont {Kravtsov}\ \emph {et~al.}(2020)\citenamefont
  {Kravtsov}, \citenamefont {Khestanova}, \citenamefont {Benimetskiy},
  \citenamefont {Ivanova}, \citenamefont {Samusev}, \citenamefont {Sinev},
  \citenamefont {Pidgayko}, \citenamefont {Mozharov}, \citenamefont {Mukhin},
  \citenamefont {Lozhkin}, \citenamefont {Kapitonov}, \citenamefont {Brichkin},
  \citenamefont {Kulakovskii}, \citenamefont {Shelykh}, \citenamefont
  {Tartakovskii}, \citenamefont {Walker}, \citenamefont {Skolnick},
  \citenamefont {Krizhanovskii},\ and\ \citenamefont {Iorsh}}]{Kravtsov2020}%
  \BibitemOpen
  \bibfield  {author} {\bibinfo {author} {\bibfnamefont {V.}~\bibnamefont
  {Kravtsov}}, \bibinfo {author} {\bibfnamefont {E.}~\bibnamefont
  {Khestanova}}, \bibinfo {author} {\bibfnamefont {F.~A.}\ \bibnamefont
  {Benimetskiy}}, \bibinfo {author} {\bibfnamefont {T.}~\bibnamefont
  {Ivanova}}, \bibinfo {author} {\bibfnamefont {A.~K.}\ \bibnamefont
  {Samusev}}, \bibinfo {author} {\bibfnamefont {I.~S.}\ \bibnamefont {Sinev}},
  \bibinfo {author} {\bibfnamefont {D.}~\bibnamefont {Pidgayko}}, \bibinfo
  {author} {\bibfnamefont {A.~M.}\ \bibnamefont {Mozharov}}, \bibinfo {author}
  {\bibfnamefont {I.~S.}\ \bibnamefont {Mukhin}}, \bibinfo {author}
  {\bibfnamefont {M.~S.}\ \bibnamefont {Lozhkin}}, \bibinfo {author}
  {\bibfnamefont {Y.~V.}\ \bibnamefont {Kapitonov}}, \bibinfo {author}
  {\bibfnamefont {A.~S.}\ \bibnamefont {Brichkin}}, \bibinfo {author}
  {\bibfnamefont {V.~D.}\ \bibnamefont {Kulakovskii}}, \bibinfo {author}
  {\bibfnamefont {I.~A.}\ \bibnamefont {Shelykh}}, \bibinfo {author}
  {\bibfnamefont {A.~I.}\ \bibnamefont {Tartakovskii}}, \bibinfo {author}
  {\bibfnamefont {P.~M.}\ \bibnamefont {Walker}}, \bibinfo {author}
  {\bibfnamefont {M.~S.}\ \bibnamefont {Skolnick}}, \bibinfo {author}
  {\bibfnamefont {D.~N.}\ \bibnamefont {Krizhanovskii}}, \ and\ \bibinfo
  {author} {\bibfnamefont {I.~V.}\ \bibnamefont {Iorsh}},\ }\bibfield  {title}
  {\enquote {\bibinfo {title} {Nonlinear polaritons in a monolayer
  semiconductor coupled to optical bound states in the continuum},}\ }\href
  {\doibase 10.1038/s41377-020-0286-z} {\bibfield  {journal} {\bibinfo
  {journal} {Light: Science {\&} Applications}\ }\textbf {\bibinfo {volume}
  {9}} (\bibinfo {year} {2020}),\ 10.1038/s41377-020-0286-z}\BibitemShut
  {NoStop}%
\bibitem [{\citenamefont {Dang}\ \emph {et~al.}(2020)\citenamefont {Dang},
  \citenamefont {Gerace}, \citenamefont {Drouard}, \citenamefont
  {Tripp{\'{e}}-Allard}, \citenamefont {L{\'{e}}d{\'{e}}e}, \citenamefont
  {Mazurczyk}, \citenamefont {Deleporte}, \citenamefont {Seassal},\ and\
  \citenamefont {Nguyen}}]{Dang2020}%
  \BibitemOpen
  \bibfield  {author} {\bibinfo {author} {\bibfnamefont {N.~H.~M.}\
  \bibnamefont {Dang}}, \bibinfo {author} {\bibfnamefont {D.}~\bibnamefont
  {Gerace}}, \bibinfo {author} {\bibfnamefont {E.}~\bibnamefont {Drouard}},
  \bibinfo {author} {\bibfnamefont {G.}~\bibnamefont {Tripp{\'{e}}-Allard}},
  \bibinfo {author} {\bibfnamefont {F.}~\bibnamefont {L{\'{e}}d{\'{e}}e}},
  \bibinfo {author} {\bibfnamefont {R.}~\bibnamefont {Mazurczyk}}, \bibinfo
  {author} {\bibfnamefont {E.}~\bibnamefont {Deleporte}}, \bibinfo {author}
  {\bibfnamefont {C.}~\bibnamefont {Seassal}}, \ and\ \bibinfo {author}
  {\bibfnamefont {H.~S.}\ \bibnamefont {Nguyen}},\ }\bibfield  {title}
  {\enquote {\bibinfo {title} {Tailoring dispersion of room-temperature
  exciton-polaritons with perovskite-based subwavelength metasurfaces},}\
  }\href {\doibase 10.1021/acs.nanolett.0c00125} {\bibfield  {journal}
  {\bibinfo  {journal} {Nano Letters}\ }\textbf {\bibinfo {volume} {20}},\
  \bibinfo {pages} {2113--2119} (\bibinfo {year} {2020})}\BibitemShut {NoStop}%
\bibitem [{\citenamefont {Liu}\ \emph {et~al.}(2020)\citenamefont {Liu},
  \citenamefont {Ji}, \citenamefont {Wang}, \citenamefont {Modi}, \citenamefont
  {Hwang}, \citenamefont {Zheng}, \citenamefont {Sorger}, \citenamefont {Pan},\
  and\ \citenamefont {Agarwal}}]{Liu2020}%
  \BibitemOpen
  \bibfield  {author} {\bibinfo {author} {\bibfnamefont {W.}~\bibnamefont
  {Liu}}, \bibinfo {author} {\bibfnamefont {Z.}~\bibnamefont {Ji}}, \bibinfo
  {author} {\bibfnamefont {Y.}~\bibnamefont {Wang}}, \bibinfo {author}
  {\bibfnamefont {G.}~\bibnamefont {Modi}}, \bibinfo {author} {\bibfnamefont
  {M.}~\bibnamefont {Hwang}}, \bibinfo {author} {\bibfnamefont
  {B.}~\bibnamefont {Zheng}}, \bibinfo {author} {\bibfnamefont {V.~J.}\
  \bibnamefont {Sorger}}, \bibinfo {author} {\bibfnamefont {A.}~\bibnamefont
  {Pan}}, \ and\ \bibinfo {author} {\bibfnamefont {R.}~\bibnamefont
  {Agarwal}},\ }\bibfield  {title} {\enquote {\bibinfo {title} {Generation of
  helical topological exciton-polaritons},}\ }\href {\doibase
  10.1126/science.abc4975} {\bibfield  {journal} {\bibinfo  {journal}
  {Science}\ }\textbf {\bibinfo {volume} {370}},\ \bibinfo {pages} {600--604}
  (\bibinfo {year} {2020})}\BibitemShut {NoStop}%
\bibitem [{\citenamefont {Karzig}\ \emph {et~al.}(2015)\citenamefont {Karzig},
  \citenamefont {Bardyn}, \citenamefont {Lindner},\ and\ \citenamefont
  {Refael}}]{PhysRevX.5.031001}%
  \BibitemOpen
  \bibfield  {author} {\bibinfo {author} {\bibfnamefont {T.}~\bibnamefont
  {Karzig}}, \bibinfo {author} {\bibfnamefont {C.-E.}\ \bibnamefont {Bardyn}},
  \bibinfo {author} {\bibfnamefont {N.~H.}\ \bibnamefont {Lindner}}, \ and\
  \bibinfo {author} {\bibfnamefont {G.}~\bibnamefont {Refael}},\ }\bibfield
  {title} {\enquote {\bibinfo {title} {Topological polaritons},}\ }\href
  {\doibase 10.1103/PhysRevX.5.031001} {\bibfield  {journal} {\bibinfo
  {journal} {Phys. Rev. X}\ }\textbf {\bibinfo {volume} {5}},\ \bibinfo {pages}
  {031001} (\bibinfo {year} {2015})}\BibitemShut {NoStop}%
\bibitem [{\citenamefont {Sun}\ \emph {et~al.}(2017)\citenamefont {Sun},
  \citenamefont {Wen}, \citenamefont {Yoon}, \citenamefont {Liu}, \citenamefont
  {Steger}, \citenamefont {Pfeiffer}, \citenamefont {West}, \citenamefont
  {Snoke},\ and\ \citenamefont {Nelson}}]{Sun2017}%
  \BibitemOpen
  \bibfield  {author} {\bibinfo {author} {\bibfnamefont {Y.}~\bibnamefont
  {Sun}}, \bibinfo {author} {\bibfnamefont {P.}~\bibnamefont {Wen}}, \bibinfo
  {author} {\bibfnamefont {Y.}~\bibnamefont {Yoon}}, \bibinfo {author}
  {\bibfnamefont {G.}~\bibnamefont {Liu}}, \bibinfo {author} {\bibfnamefont
  {M.}~\bibnamefont {Steger}}, \bibinfo {author} {\bibfnamefont {L.~N.}\
  \bibnamefont {Pfeiffer}}, \bibinfo {author} {\bibfnamefont {K.}~\bibnamefont
  {West}}, \bibinfo {author} {\bibfnamefont {D.~W.}\ \bibnamefont {Snoke}}, \
  and\ \bibinfo {author} {\bibfnamefont {K.~A.}\ \bibnamefont {Nelson}},\
  }\bibfield  {title} {\enquote {\bibinfo {title} {Bose-einstein condensation
  of long-lifetime polaritons in thermal equilibrium},}\ }\href {\doibase
  10.1103/physrevlett.118.016602} {\bibfield  {journal} {\bibinfo  {journal}
  {Physical Review Letters}\ }\textbf {\bibinfo {volume} {118}},\ \bibinfo
  {pages} {016602} (\bibinfo {year} {2017})}\BibitemShut {NoStop}%
\bibitem [{\citenamefont {Joannopoulos}\ \emph {et~al.}(2008)\citenamefont
  {Joannopoulos}, \citenamefont {Johnson}, \citenamefont {Winn},\ and\
  \citenamefont {Meade}}]{PhotonicCrystalsBook}%
  \BibitemOpen
  \bibfield  {author} {\bibinfo {author} {\bibfnamefont {J.~D.}\ \bibnamefont
  {Joannopoulos}}, \bibinfo {author} {\bibfnamefont {S.~G.}\ \bibnamefont
  {Johnson}}, \bibinfo {author} {\bibfnamefont {J.~N.}\ \bibnamefont {Winn}}, \
  and\ \bibinfo {author} {\bibfnamefont {R.~D.}\ \bibnamefont {Meade}},\ }\href
  {https://press.princeton.edu/books/hardcover/9780691124568/photonic-crystals}
  {\emph {\bibinfo {title} {Photonic Crystals: Molding the Flow of Light}}},\
  \bibinfo {edition} {2nd}\ ed.\ (\bibinfo  {publisher} {Princeton University
  Press},\ \bibinfo {address} {Princeton, NJ},\ \bibinfo {year}
  {2008})\BibitemShut {NoStop}%
\bibitem [{\citenamefont {Rahimi-Iman}(2020)}]{Rahimi-Iman2020}%
  \BibitemOpen
  \bibfield  {author} {\bibinfo {author} {\bibfnamefont {A.}~\bibnamefont
  {Rahimi-Iman}},\ }\enquote {\bibinfo {title} {Polaritons in external
  fields},}\ in\ \href {\doibase 10.1007/978-3-030-39333-5_9} {\emph {\bibinfo
  {booktitle} {Polariton Physics: From Dynamic Bose--Einstein Condensates in
  Strongly‐Coupled Light--Matter Systems to Polariton Lasers}}}\ (\bibinfo
  {publisher} {Springer International Publishing},\ \bibinfo {address} {Cham},\
  \bibinfo {year} {2020})\ pp.\ \bibinfo {pages} {241--262}\BibitemShut
  {NoStop}%
\bibitem [{\citenamefont {Traynor}, \citenamefont {Harley},\ and\ \citenamefont
  {Warburton}(1995)}]{PhysRevB.51.7361}%
  \BibitemOpen
  \bibfield  {author} {\bibinfo {author} {\bibfnamefont {N.~J.}\ \bibnamefont
  {Traynor}}, \bibinfo {author} {\bibfnamefont {R.~T.}\ \bibnamefont {Harley}},
  \ and\ \bibinfo {author} {\bibfnamefont {R.~J.}\ \bibnamefont {Warburton}},\
  }\bibfield  {title} {\enquote {\bibinfo {title} {Zeeman splitting and g
  factor of heavy-hole excitons in {In$_{x}$Ga$_{1-x}$As/GaAs} quantum
  wells},}\ }\href {\doibase 10.1103/PhysRevB.51.7361} {\bibfield  {journal}
  {\bibinfo  {journal} {Phys. Rev. B}\ }\textbf {\bibinfo {volume} {51}},\
  \bibinfo {pages} {7361--7364} (\bibinfo {year} {1995})}\BibitemShut {NoStop}%
\bibitem [{\citenamefont {Mirek}\ \emph {et~al.}(2017)\citenamefont {Mirek},
  \citenamefont {Kr\'ol}, \citenamefont {Lekenta}, \citenamefont {Rousset},
  \citenamefont {Nawrocki}, \citenamefont {Kulczykowski}, \citenamefont
  {Matuszewski}, \citenamefont {Szczytko}, \citenamefont {Pacuski},\ and\
  \citenamefont {Pi\ifmmode~\mbox{\k{e}}\else
  \k{e}\fi{}tka}}]{PhysRevB.95.085429}%
  \BibitemOpen
  \bibfield  {author} {\bibinfo {author} {\bibfnamefont {R.}~\bibnamefont
  {Mirek}}, \bibinfo {author} {\bibfnamefont {M.}~\bibnamefont {Kr\'ol}},
  \bibinfo {author} {\bibfnamefont {K.}~\bibnamefont {Lekenta}}, \bibinfo
  {author} {\bibfnamefont {J.-G.}\ \bibnamefont {Rousset}}, \bibinfo {author}
  {\bibfnamefont {M.}~\bibnamefont {Nawrocki}}, \bibinfo {author}
  {\bibfnamefont {M.}~\bibnamefont {Kulczykowski}}, \bibinfo {author}
  {\bibfnamefont {M.}~\bibnamefont {Matuszewski}}, \bibinfo {author}
  {\bibfnamefont {J.}~\bibnamefont {Szczytko}}, \bibinfo {author}
  {\bibfnamefont {W.}~\bibnamefont {Pacuski}}, \ and\ \bibinfo {author}
  {\bibfnamefont {B.}~\bibnamefont {Pi\ifmmode~\mbox{\k{e}}\else
  \k{e}\fi{}tka}},\ }\bibfield  {title} {\enquote {\bibinfo {title} {Angular
  dependence of giant {Zeeman} effect for semimagnetic cavity polaritons},}\
  }\href {\doibase 10.1103/PhysRevB.95.085429} {\bibfield  {journal} {\bibinfo
  {journal} {Phys. Rev. B}\ }\textbf {\bibinfo {volume} {95}},\ \bibinfo
  {pages} {085429} (\bibinfo {year} {2017})}\BibitemShut {NoStop}%
\bibitem [{Note1()}]{Note1}%
  \BibitemOpen
  \bibinfo {note} {A. I. Tartakovskii, private communication
  (unpublished)}\BibitemShut {NoStop}%
\bibitem [{\citenamefont {Jacqmin}\ \emph {et~al.}(2014)\citenamefont
  {Jacqmin}, \citenamefont {Carusotto}, \citenamefont {Sagnes}, \citenamefont
  {Abbarchi}, \citenamefont {Solnyshkov}, \citenamefont {Malpuech},
  \citenamefont {Galopin}, \citenamefont {Lema\^{\i}tre}, \citenamefont
  {Bloch},\ and\ \citenamefont {Amo}}]{PhysRevLett.112.116402}%
  \BibitemOpen
  \bibfield  {author} {\bibinfo {author} {\bibfnamefont {T.}~\bibnamefont
  {Jacqmin}}, \bibinfo {author} {\bibfnamefont {I.}~\bibnamefont {Carusotto}},
  \bibinfo {author} {\bibfnamefont {I.}~\bibnamefont {Sagnes}}, \bibinfo
  {author} {\bibfnamefont {M.}~\bibnamefont {Abbarchi}}, \bibinfo {author}
  {\bibfnamefont {D.~D.}\ \bibnamefont {Solnyshkov}}, \bibinfo {author}
  {\bibfnamefont {G.}~\bibnamefont {Malpuech}}, \bibinfo {author}
  {\bibfnamefont {E.}~\bibnamefont {Galopin}}, \bibinfo {author} {\bibfnamefont
  {A.}~\bibnamefont {Lema\^{\i}tre}}, \bibinfo {author} {\bibfnamefont
  {J.}~\bibnamefont {Bloch}}, \ and\ \bibinfo {author} {\bibfnamefont
  {A.}~\bibnamefont {Amo}},\ }\bibfield  {title} {\enquote {\bibinfo {title}
  {Direct observation of {Dirac} cones and a flatband in a honeycomb lattice
  for polaritons},}\ }\href {\doibase 10.1103/PhysRevLett.112.116402}
  {\bibfield  {journal} {\bibinfo  {journal} {Phys. Rev. Lett.}\ }\textbf
  {\bibinfo {volume} {112}},\ \bibinfo {pages} {116402} (\bibinfo {year}
  {2014})}\BibitemShut {NoStop}%
\bibitem [{\citenamefont {Whittaker}\ \emph {et~al.}(2018)\citenamefont
  {Whittaker}, \citenamefont {Cancellieri}, \citenamefont {Walker},
  \citenamefont {Gulevich}, \citenamefont {Schomerus}, \citenamefont
  {Vaitiekus}, \citenamefont {Royall}, \citenamefont {Whittaker}, \citenamefont
  {Clarke}, \citenamefont {Iorsh}, \citenamefont {Shelykh}, \citenamefont
  {Skolnick},\ and\ \citenamefont {Krizhanovskii}}]{PhysRevLett.120.097401}%
  \BibitemOpen
  \bibfield  {author} {\bibinfo {author} {\bibfnamefont {C.~E.}\ \bibnamefont
  {Whittaker}}, \bibinfo {author} {\bibfnamefont {E.}~\bibnamefont
  {Cancellieri}}, \bibinfo {author} {\bibfnamefont {P.~M.}\ \bibnamefont
  {Walker}}, \bibinfo {author} {\bibfnamefont {D.~R.}\ \bibnamefont
  {Gulevich}}, \bibinfo {author} {\bibfnamefont {H.}~\bibnamefont {Schomerus}},
  \bibinfo {author} {\bibfnamefont {D.}~\bibnamefont {Vaitiekus}}, \bibinfo
  {author} {\bibfnamefont {B.}~\bibnamefont {Royall}}, \bibinfo {author}
  {\bibfnamefont {D.~M.}\ \bibnamefont {Whittaker}}, \bibinfo {author}
  {\bibfnamefont {E.}~\bibnamefont {Clarke}}, \bibinfo {author} {\bibfnamefont
  {I.~V.}\ \bibnamefont {Iorsh}}, \bibinfo {author} {\bibfnamefont {I.~A.}\
  \bibnamefont {Shelykh}}, \bibinfo {author} {\bibfnamefont {M.~S.}\
  \bibnamefont {Skolnick}}, \ and\ \bibinfo {author} {\bibfnamefont {D.~N.}\
  \bibnamefont {Krizhanovskii}},\ }\bibfield  {title} {\enquote {\bibinfo
  {title} {Exciton polaritons in a two-dimensional lieb lattice with spin-orbit
  coupling},}\ }\href {\doibase 10.1103/PhysRevLett.120.097401} {\bibfield
  {journal} {\bibinfo  {journal} {Phys. Rev. Lett.}\ }\textbf {\bibinfo
  {volume} {120}},\ \bibinfo {pages} {097401} (\bibinfo {year}
  {2018})}\BibitemShut {NoStop}%
\bibitem [{\citenamefont {Jayaprakash}\ \emph {et~al.}(2020)\citenamefont
  {Jayaprakash}, \citenamefont {Whittaker}, \citenamefont {Georgiou},
  \citenamefont {Game}, \citenamefont {McGhee}, \citenamefont {Coles},\ and\
  \citenamefont {Lidzey}}]{Jayaprakash2020}%
  \BibitemOpen
  \bibfield  {author} {\bibinfo {author} {\bibfnamefont {R.}~\bibnamefont
  {Jayaprakash}}, \bibinfo {author} {\bibfnamefont {C.~E.}\ \bibnamefont
  {Whittaker}}, \bibinfo {author} {\bibfnamefont {K.}~\bibnamefont {Georgiou}},
  \bibinfo {author} {\bibfnamefont {O.~S.}\ \bibnamefont {Game}}, \bibinfo
  {author} {\bibfnamefont {K.~E.}\ \bibnamefont {McGhee}}, \bibinfo {author}
  {\bibfnamefont {D.~M.}\ \bibnamefont {Coles}}, \ and\ \bibinfo {author}
  {\bibfnamefont {D.~G.}\ \bibnamefont {Lidzey}},\ }\bibfield  {title}
  {\enquote {\bibinfo {title} {Two-dimensional organic-exciton polariton
  lattice fabricated using laser patterning},}\ }\href {\doibase
  10.1021/acsphotonics.0c00867} {\bibfield  {journal} {\bibinfo  {journal} {ACS
  Photonics}\ }\textbf {\bibinfo {volume} {7}},\ \bibinfo {pages} {2273--2281}
  (\bibinfo {year} {2020})}\BibitemShut {NoStop}%
\end{thebibliography}%



\end{document}